\documentclass[aps,twocolumn,showpacs,floatfix]{revtex4}

\usepackage{amsmath}
\usepackage{latexsym}
\usepackage{float}
\usepackage{amssymb}
\usepackage{graphicx}

\newcommand{\figwidth}{0.95\columnwidth}
\newcommand{\eq}[1]{Eq.(\ref{#1})}
\newcommand{\sect}[1]{Section~\ref{#1}}
\newcommand{\tab}[1]{Table~\ref{#1}}
\newcommand{\fig}[1]{Fig.~\ref{#1}}
\newcommand{\avg}[1]{ {\langle #1 \rangle} }
\newcommand{\olcite}[1]{Ref.~\onlinecite{#1}}

\newcommand{\sigmac}{\sigma_{\rm c}}
\newcommand{\sigmap}{\sigma_{\rm p}}
\newcommand{\Nc}{N_{\rm c}}
\newcommand{\zc}{z_{\rm c}}
\newcommand{\zp}{z_{\rm p}}
\newcommand{\etac}{\eta_{\rm c} }
\newcommand{\etaccr}{\eta_{\rm c,cr} }
\newcommand{\etacv}{\eta_{\rm c}^{\rm v}}
\newcommand{\etacl}{\eta_{\rm c}^{\rm l}}
\newcommand{\etapr}{\eta_{\rm p}^{\rm r}}
\newcommand{\etaprcr}{\eta_{\rm p,cr}^{\rm r}}
\newcommand{\pc}{P_L(\etac|\etapr,\zc)}
\newcommand{\lx}{L_{\rm x}}
\newcommand{\ly}{L_{\rm y}}
\newcommand{\lz}{L_{\rm z}}
\newcommand{\kb}{k_{\rm B}}

\begin{document}

\title{Critical behavior of a colloid-polymer mixture 
confined between walls}

\author{R. L. C. Vink}
\affiliation{Institut Theoretische Physik II, Heinrich Heine
Universit\"at, Universit\"atsstra{\ss}e 1, 40225 D\"usseldorf, Germany}

\author{K. Binder and J. Horbach}
\affiliation{Institut f\"ur Physik, Johannes Gutenberg-Universit\"at,
Staudinger Weg 7, 55099 Mainz, Germany}

\date{\today}

\begin{abstract} 

We investigate the influence of confinement on phase separation in
colloid-polymer mixtures. To describe the particle interactions, the
colloid-polymer model of Asakura and Oosawa [J. Chem. Phys. {\bf 22}, 1255
(1954)] is used. Grand canonical Monte Carlo simulations are then applied to
this model confined between two parallel hard walls, separated by a distance
$D=5$ colloid diameters. We focus on the critical regime of the phase separation
and look for signs of crossover from three-dimensional (3D) Ising to
two-dimensional (2D) Ising universality. To extract the critical behavior,
finite size scaling techniques are used, including the recently proposed
algorithm of Kim {\it et al.} [Phys.~Rev.~Lett.~{\bf 91}, 065701 (2003)]. Our
results point to ``effective'' critical exponents that differ profoundly from 3D
Ising values, and that are already very close to 2D Ising values. In particular,
we observe that the critical exponent $\beta$ of the order parameter in the
confined system is smaller than in 3D bulk, yielding a ``flatter'' binodal. Our
results also show an increase in the critical colloid packing fraction in the
confined system with respect to the bulk. The latter seems consistent with
theoretical expectations, although subtleties due to singularities in the
critical behavior of the coexistence diameter cannot be ruled out.

\end{abstract}


\maketitle

\section{Introduction}

The current technological demand for the production of nanoscopic devices 
\cite{1,2,3,4,5} renewed the interest in understanding the phase behavior 
of fluid systems confined in pores of nanoscopic linear dimensions 
\cite{6,7}. In addition, porous materials with pore widths less than 50 nm 
are widely used in the chemical, oil and gas, food, and pharmaceutical 
industries, for applications such as mixture separation, pollution 
control, and as catalysts \cite{6,8,9,10}. However, many such applications 
rely largely on empirical knowledge, since the theory-based understanding 
of confined fluids is still rather incomplete \cite{6,11,12,13,14,15}. 
Even the basic phenomenon of ``capillary condensation'' of undersaturated 
gases in capillaries, described already in the 19$^{\rm th}$ century 
\cite{16}, still forms the subject of longstanding investigations by 
analytical theory \cite{17,18,19,20,21,22,23,24,25} and computer 
simulations \cite{20,26,27,28,29,30,31,32,33,34,35}. Regarding confined 
binary mixtures, there is a close analogy between the phase behavior of 
confined one-component fluids and the preferential adsorption of one of 
the components of the mixture to the walls. The miscibility gap of the 
mixture corresponds to the coexistence curve (or binodal) that describes 
the phase separation between liquid and gas in simple fluids, and numerous 
theoretical and simulational studies have addressed the phase behavior of 
binary mixtures in cylindrical pores or slit pores 
\cite{36,37,38,39,40,41,42,43,44,45,46,47}. Depending on the details of 
the wall-particle interactions in relation to the interactions among the 
fluid particles of a binary (AB) mixture, it is clear that there can be 
either the A-component or the B-component attracted to the walls, apart 
from the very special case of ``neutral walls'' which produce confinement 
only \cite{48, 49, 50}. Similarly, in a one-component liquid-gas system 
``capillary evaporation'' \cite{46,51,52,53} can occur for repulsive 
wall-particle interactions. While for the liquid-gas transition and the 
demixing transition of binary fluids the ``order parameter'' of the 
transition is a simple scalar, i.e.~the transition belongs to the 
``universality class'' of the Ising (lattice gas) model \cite{54}, related 
phenomena occur in systems with more complex ordering, e.g.~confined 
liquid crystals where ``capillary nematization'' may occur \cite{55}.

Understanding nanoscopic confinement of fluids consisting of small 
molecules is difficult because the lateral variation of the wall 
potential, due to wall roughness or even atomistic corrugation \cite{56} 
of the wall potential, may have drastic effects on the phase behavior of 
the confined fluid \cite{57}. In this respect, colloidal systems due to 
the mesoscopic size of the colloidal particles pose distinct advantages: 
atomistic corrugation of the confining wall potentials can safely be 
neglected, and there is a great freedom in preparing systems with suitable 
interactions \cite{58,59,60}. A particularly suitable class of systems are 
colloid-polymer mixtures, since both bulk phase behavior and the 
interfaces separating the colloid-rich and polymer-rich phases can be 
studied experimentally in detail \cite{61,62,63,64}. Furthermore, the 
Asakura-Oosawa (AO) model \cite{65,66} provides a simple theoretical 
description, which seems to capture all the salient features of such 
phase-separating colloid-polymer mixtures, and which is well suited to 
computer simulation investigations \cite{46,67,68,70}.

Therefore, we use this model again in the present paper to address the 
question: how does the critical behavior near the demixing critical point 
change due to the confinement in slit pores? While on general theoretical 
grounds one expects \cite{15,18,34,35,48,49,50} that in the ultimate 
vicinity asymptotically close to the critical point the critical exponents 
of the two-dimensional (2D) Ising model should apply \cite{71}, this 
limiting asymptotic behavior may be hard to observe, and the need arises 
to consider the crossover from three dimensional (3D) to two-dimensional 
critical behavior in such thin films \cite{48,49,50}. This crossover 
behavior in itself is a difficult but interesting problem \cite{49}. 
Previous work considering this issue has been restricted to either simple 
Ising (lattice gas) models \cite{18,35,48} or strictly symmetric polymer 
mixtures confined by neutral walls \cite{50}. The complications due to the 
asymmetry between liquid(-like) and gas(-like) phases in the bulk have not 
been considered, and also the further asymmetry arising from the 
preferential adsorption of one species to the walls has been disregarded. 
Due to this preferential adsorption, one expects near the critical point 
of the bulk the formation of wetting layers at the walls 
\cite{15,72,73,74,75} if the width of the slit pore becomes very large.

We note that previous theoretical work on capillary condensation (or 
evaporation) based on density functional theory (DFT), or other analytical 
approximations for the equation of state, inevitably implies a parabolic 
shape of the binodal near the critical value $\etaprcr$ of the polymer 
reservoir packing fraction $\etapr$, irrespective of whether one considers 
the bulk mixture or a confined system. Therefore, despite the fact that 
such theories are very powerful away from the critical point, they cannot 
be used to describe the crossover in the critical behavior due to 
confinement. Also previous Gibbs ensemble simulations of the AO model 
confined to slit pores \cite{46} did not address this issue since the 
Gibbs ensemble cannot be used to sample the critical point. In the present 
work, we therefore extend the {\it grand canonical} techniques used in our 
previous work on the critical behavior of the AO model in the bulk 
\cite{67,68} to study the critical behavior in confinement.

The outline of this paper is as follows. In \sect{sec:theory}, we briefly 
review crossover scaling relations that are expected to describe the critical 
behavior of a confined fluid. Next, we introduce the AO model and describe 
our simulation method. In \sect{sec:model}, we present our results, using 
a finite size scaling analysis of the critical properties for a very thin 
film of thickness $D = 5 \sigmac$, with $\sigmac$ the diameter of the 
hard-sphere colloids. We end with a summary and conclusion in the last 
section.

\section{Crossover scaling}
\label{sec:theory}

To define the problem of study more precisely, we recall that the width of 
the binodal, or order parameter, of a colloid-polymer mixture is given by
\begin{equation}
\label{eq:order}
  \Delta \equiv (\etacl-\etacv)/2,
\end{equation}
with $\etacl$ ($\etacv$) the colloid packing fraction in the colloidal liquid
(vapor) phase. In 3D bulk, the order parameter close to the critical point is
expected to scale as
\begin{equation} 
\label{eq1}
  \Delta(\infty) = \hat{B} t_\infty^\beta, \quad
  t_\infty \equiv \etapr/\etaprcr(\infty) - 1,
\end{equation}
where $\hat{B}$ is a (nonuniversal) critical amplitude, and $\beta 
\approx 0.326$ the (universal) critical exponent of the 3D Ising 
universality class \cite{54,76,77}. The symbol $(\infty)$ in the 
above emphasizes that $\etaprcr$ is the critical value of the polymer 
reservoir packing fraction appropriate for an infinitely thick film, 
i.e.~a bulk 3D system.

In a confined thin film of thickness $D$, however, the corresponding 
relation reads
\begin{equation} 
\label{eq2}
  \Delta(D) = \hat{B}(D) t^{\beta_2}_D, \quad
  t_D \equiv \etapr / \etaprcr(D) - 1.
\end{equation}
The critical polymer reservoir packing fraction is thus shifted from its bulk
value $\etaprcr(\infty)$ to a new value $\etaprcr(D)$. In addition, the critical
amplitude $\hat{B}(D)$ depends on the film thickness $D$, and the critical
exponent takes the value of the 2D Ising universality class $\beta_2=1/8$
\cite{54,71}. However, as $D$ gets large, the validity of \eq{eq2} is expected
to be observable only in an extremely narrow region around $t_D=0$. This is
recognized when one formulates the appropriate crossover scaling description
\cite{18,35,48,50,78}
\begin{equation} 
\label{eq3}
 \Delta = D^{-\beta/\nu} F (D^{1/\nu} t_\infty), 
\end{equation}
where $\nu$ is the critical exponent of the correlation length for the 3D Ising
universality class \cite{54,76,77}, and $F(X)$ a crossover scaling
function with $X \equiv D^{1/ \nu}t_\infty$. \eq{eq3} may qualitatively be
interpreted using the finite size scaling principle \cite{79,80,81,82} in which
the film thickness $D$ scales with the correlation length $\xi= \hat{\xi}
t^{-\nu}_\infty$, where $\hat{\xi}$ is another critical amplitude. To recover
\eq{eq1} from \eq{eq3}, one notes that the scaling function $F(X)$ must behave
as $F(X) \propto X^\beta$ for $X \rightarrow \infty$.  At a fixed small value of
$t_\infty$, the $D$-dependence then cancels out from the equation, as it should.
On the other hand, \eq{eq2} may also be recovered from \eq{eq3}, by postulating
that for small $X$ a singularity occurs when $X$ approaches $X_{\rm crit}$,
namely
\begin{equation} 
\label{eq4}
  F(X) = \hat{f} (X-X_{\rm crit})^{\beta_2}, \quad X-X_{\rm
  crit} \ll 1,
\end{equation}
with $\hat{f}$ another non-universal amplitude. This phenomenological 
assumption implies a scaling relation for the shift of the critical 
value of the polymer reservoir packing fraction
\begin{equation} 
\label{eq5}
  X_{\rm crit} = D^{1/\nu} t^{\rm crit}_\infty \Rightarrow t^{\rm
  crit}_\infty = X_{\rm crit} D^{-1/\nu}.
\end{equation}

Another scaling relation is implied for the critical amplitude
$\hat{B} (D)$, namely
\begin{equation} \label{eq6}
\hat{B} (D) =\hat{f} D^{(\beta_2 - \beta)/\nu}.
\end{equation}
It is clear that the crossover between both power laws, \eq{eq1} and \eq{eq2},
then also should occur when $X$ is of order unity, which implies very small
values of $t_\infty$ already when $D$ is large. The region of $t_\infty -
t_{\infty, {\rm crit}}$ where \eq{eq2} and \eq{eq4} then hold is extremely
small. Moreover, the general experience with problems of this kind is that a
crossover never occurs abruptly \cite{77, luijten.binder:1998,
anisimov.luijten.ea:1999, luijten.meyer:2000}, but rather spans several decades
of the corresponding crossover scaling variable ($X-X_{\rm crit}$ in our case). 
In fact, if not a large enough range of this crossover scaling variable is
accessible, one will instead observe a power law with ``effective exponents''
and ``effective critical amplitudes'',
\begin{equation} 
\label{eq7}
  \Delta (D) \approx \hat{B}_{\rm eff} t^{\beta{\rm eff}}_D,
\end{equation}
where $\beta_2 < \beta_{\rm eff} < \beta$ \cite{50}. The effective exponents 
do not have a fundamental deep meaning, of course, since their values depend 
on the range of $t_D$ that is used for the analysis in terms of \eq{eq7}, and 
hence are not really defined unambiguously.

\section{Model and Simulation method}
\label{sec:model}

We consider a mixture of hard-sphere colloids with diameter $\sigmac$ and
effective polymer spheres with diameter of gyration $\sigmap$ in the grand
canonical ensemble. Throughout this work, the colloid diameter $\sigmac$ is
taken to be the unit of length. In the grand canonical ensemble, the volume $V$
and the respective fugacities, $\zc$ and $\zp$, of colloids and polymers are
fixed, while the number of particles in the system fluctuates. Following
convention, the polymer fugacity is expressed in terms of a related quantity
called the polymer reservoir packing fraction $\etapr = \pi \zp \sigmap^3 / 6$.
We also introduce the colloid packing fraction $\etac = \pi \sigmac^3 \Nc /
(6V)$, with $\Nc$ the number of colloids in the system.  The particles interact
via potentials that were originally proposed by Asakura and Oosawa \cite{65}
(AO) and later, independently, also by Vrij \cite{66}. In this description, the
so-called AO model, hard-sphere interactions are assumed between colloid-colloid
and colloid-polymer pairs, while polymer-polymer pairs can interpenetrate
freely. The interactions are thus strictly athermal such that the temperature
plays no role. Instead, in the AO model, the analogue of (inverse) temperature
is played by the polymer reservoir packing fraction $\etapr$. As is well known,
at the coexistence colloid fugacity and for sufficiently large
colloid-to-polymer size ratios $q \equiv \sigmap / \sigmac$, the AO model 
exhibits a phase separation into a colloid-rich phase (the colloidal liquid) and
colloid-poor phase (the colloidal vapor), provided $\etapr$ exceeds a critical
value. Grand canonical Monte Carlo simulations are well suited to study this
transition, and when combined with finite size scaling techniques these
simulations also allow for investigations close to the critical point. Recently,
this approach was applied to the bulk AO model, i.e.\ in the absence of walls,
and the critical point was determined for $q=0.8$, as well as the universality
class, which was shown to be that of the 3D Ising model \cite{67,68}.

In this work, grand canonical Monte Carlo simulations are used to study the AO
model in confinement. To this end, we use a simulation box of dimensions $\lx
\times \ly \times \lz$, with $\lx=\ly=L$ and $\lz=D$; the system volume thus
equals $V=D L^2$. To capture the effect of confinement, we implement a so-called
``sandwich'' or ``thin film'' geometry. Here, periodic boundary conditions are
applied in the $x$ and $y$ directions, while in the remaining $z$ direction we
place two parallel walls: one in the $z=0$ plane, and in the $z=\lz$ plane. This
closely resembles \olcite{46} where capillary condensation and evaporation of
the AO model are investigated. Compared to the bulk AO model, one additional
parameter is thus introduced, namely the film thickness $D$. For a film with
thickness $D$, the thermodynamic limit is defined as the limit where the lateral
dimensions $\lx=\ly=L$ of the film are taken to infinity. Throughout this work,
the walls are taken to be hard, i.e.\ colloid-wall and polymer-wall overlaps are
strictly forbidden.  Note that this implies a strong attraction between the
colloids and the walls due to the depletion effect \cite{46}. The simulation
method of \olcite{67} is then applied to the confined system; the main
ingredients are a grand canonical cluster move \cite{67} and a reweighting
scheme \cite{virnau.muller:2004}.

\section{Results}

\begin{figure}
\begin{center}
\includegraphics[width=\figwidth,clip]{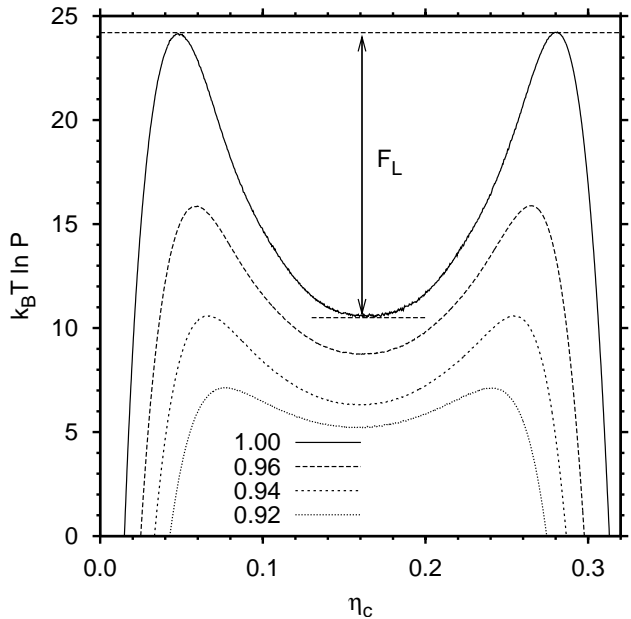}

\caption{\label{prob} Coexistence distributions of the confined AO model 
with colloid-to-polymer size ratio $q=0.8$, lateral dimension $L=15$, and 
film thickness $D=5$ for several values of $\etapr$ as indicated. Note 
that we have plotted the {\it natural logarithm} of $\pc$, multiplied by 
$\kb T$. In the above distributions, the colloid fugacity is tuned in 
order to obey the equal-weight prescription \cite{muller:1995}. The 
barrier $F_L$ in the distribution corresponding to $\etapr=1.0$ marks the 
average height of the peaks with respect to the minimum in between the 
peaks.}

\end{center}
\end{figure}

\subsection{Binodal}

For $q=0.8$ and $D=5$, the grand canonical distribution $\pc$ is measured, 
defined as the probability of observing a system with colloid packing 
fraction $\etac$, at ``inverse temperature'' $\etapr$ and colloid fugacity 
$\zc$. There will generally be finite size effects in the lateral 
dimensions $\lx=\ly=L$ of the simulation box, denoted by the subscript 
$L$. Phase coexistence is established by tuning $\zc$ such that $\pc$ 
becomes bimodal with two peaks of equal area \cite{muller:1995}. The 
respective packing fractions $\etacv$ and $\etacl$, of the colloidal vapor 
and liquid phase, are obtained from the average peak positions. Typical 
distributions are shown in \fig{prob}, plotted as $\kb T \ln \pc$, with 
$\kb$ the Boltzmann constant and $T$ the temperature. In this way, the 
distributions correspond to {\it minus} the free energy of the system. The 
height $F_L$ of the peaks in $\kb T \ln \pc$, measured with respect to the 
minimum in between the peaks (arrow in \fig{prob}), thus reflects the free 
energy barrier separating the colloidal vapor from the liquid phase 
\cite{binder:1982}. In the two-phase region away from the critical point, 
the peaks in $\pc$ are well separated and the barrier $F_L$ will be large. 
Upon approach of the critical point, by lowering $\etapr$, the peaks move 
closer together and the corresponding barrier $F_L$ decreases profoundly.

\begin{figure}
\begin{center}
\includegraphics[width=\figwidth,clip]{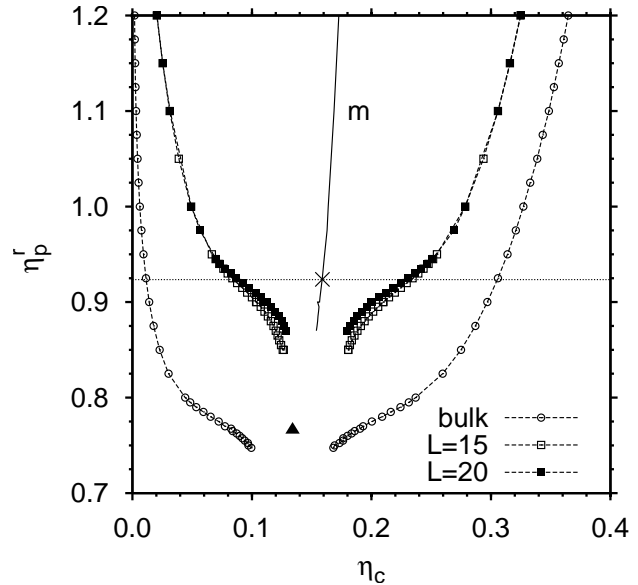}
\caption{\label{phasediag} Binodal of the AO model with $q=0.8$ in bulk 
and confinement. Open circles show the bulk binodal, where the black triangle 
marks the corresponding critical point $(\etac=0.134; \, \etapr=0.766)$ 
obtained using finite size scaling \cite{67,68}. Open and closed squares 
show the binodal of the confined system with film thickness $D=5$ for two 
lateral dimensions $L$. The horizontal line marks the critical polymer 
reservoir packing fraction $\etaprcr$ of the confined system in the limit $L 
\to \infty$ obtained using the cumulant intersection method, see 
\fig{cumulant}. Lines connecting the points serve to guide the eye. Curve $m$ 
shows the coexistence diameter $(\etacl+\etacv)/2$ of the confined system 
with $L=20$.}
\end{center}
\end{figure}

To obtain the binodal, $\pc$ is measured for a range of $\etapr$ and the 
average peak positions are recorded. The result is shown in 
\fig{phasediag}. For comparison, the binodal of the bulk AO model is also 
shown, together with the bulk critical point taken from previous work 
\cite{67,68}. Note the familiar finite-size rounding in the simulation 
data close to the critical point (to describe the binodal correctly near 
the critical point requires finite size scaling, which we postpone to 
\fig{coex_curve}). The phase diagrams in \fig{phasediag} reveal the 
typical behavior of a fluid confined between two plates that undergoes
a demixing transition: The critical point shows a significant shift
which is accompanied by an inward shift of the binodal with respect to
the bulk \cite{18}.

\begin{figure}
\begin{center}
\includegraphics[width=\figwidth,clip]{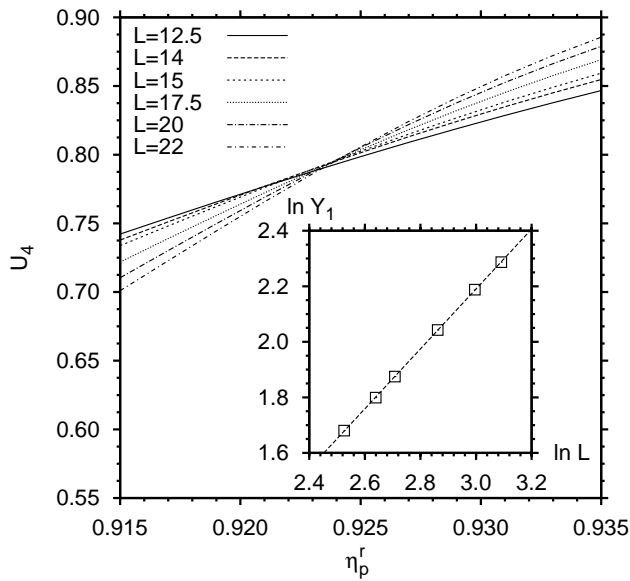}

\caption{\label{cumulant} Cumulant analysis of the confined AO model with 
$q=0.8$ and film thickness $D=5$. Shown is the fourth order cumulant $U_4$ 
as function of $\etapr$ for several lateral dimensions $L$. The 
intersection point yields an estimate of $\etaprcr$. The inset shows the 
cumulant slope $Y_1$ at the intersection point as function of the lateral 
dimension $L$.} 

\end{center} 
\end{figure}

\subsection{Cumulant analysis}
\label{sec:cum}

To obtain the critical polymer reservoir packing fraction $\etaprcr$ of the 
confined system, the fourth order cumulant $U_4 = \avg{m^2}^2 / \avg{m^4}$ 
is measured, with $m = \etac - \avg{\etac}$, as function of $\etapr$ for 
various lateral dimensions $L$. The cumulant is obtained by taking 
appropriate moments of the distribution $\pc$. For example, the average 
colloid packing fraction may be written as $\avg{\etac}(L,\etapr,\zc) = 
\int_0^\infty \etac \pc \, d \etac$, and similarly for the $p$-th order 
moment $\avg{m^p}(L,\etapr,\zc) = \int_0^\infty \left[ \etac - \avg{\etac} 
\right]^p \pc \, d \etac$. Note that the outcome will generally depend on 
$\etapr$, the colloid fugacity $\zc$, and the lateral system size $L$. 

\fig{cumulant} shows the cumulant as function of $\etapr$ for several 
system sizes $L$. We emphasize that the measurements were taken using the 
colloid fugacity at which the equal-weight prescription \cite{muller:1995} 
was obeyed. At the critical point, the cumulant becomes system-size 
independent \cite{binder:1981}. The intersections in \fig{cumulant} thus 
provide an estimate of the critical polymer reservoir packing fraction. We 
obtain $\etaprcr = 0.9238 \pm 0.0010$, where the error reflects the 
scatter in the various intersection points. This estimate is also shown in 
\fig{phasediag} (horizontal line). Defining the coexistence diameter as 
$(\etacl+\etacv)/2$ and ignoring finite size effects in this quantity for 
the moment, the intersection of the horizontal line with curve $m$ (marked 
with a cross in \fig{phasediag}) yields an estimate of the critical 
colloid packing fraction $\etaccr \approx 0.159$. Compared to the bulk 
system, the critical colloid packing fraction has shifted to a slightly 
larger value. 

\begin{table}[b]

\caption{\label{tab:prop} Critical exponents of the order parameter 
$(\beta)$, correlation length $(\nu)$, and specific heat $(\alpha)$ for 
the two-dimensional (2D) and three-dimensional (3D) Ising model, as well 
as the mean-field values (MF). Also listed is the value of the fourth 
order cumulant $(U_4)$ at criticality for the 2D and 3D Ising model.}

\begin{ruledtabular}
\begin{tabular}{ccccc}
   & $\beta$ & $\nu$ & $\alpha$ & $U_4$ \\ \hline
2D &  1/8    & 1     & 0        & 0.856 \cite{kamieniarz.blote:1993} \\
3D &  0.326 \cite{fisher.zinn:1998}  & 0.630 \cite{fisher.zinn:1998} & 
0.109 \cite{fisher.zinn:1998}  & 0.629\footnote{obtained from the 3D 
Ising universal fixed point distribution of \olcite{wilding:1996}.} \\ 
MF & 1/2 & 1/2 & 0 & -- \\
\end{tabular}
\end{ruledtabular}

\end{table}

The cumulant plot also provides evidence for the crossover scenario 
discussed in the introduction. From \fig{cumulant}, we obtain $U_4 \approx 
0.795$ at the critical point, which is between the 2D and 3D Ising values, 
see \tab{tab:prop}. In the limit $L \to \infty$, $U_4$ is expected to 
approach the 2D Ising value. However, in the (still moderate) system sizes 
accessible in our simulations, ``effective'' critical behavior is 
observed instead, with properties between those of the 2D and 3D Ising 
universality classes. Additional confirmation of the crossover may be obtained from an 
analysis of the cumulant slope $Y_1 \equiv d U_4 / d \etapr$ evaluated at 
the critical value of $\etapr$. It is expected that $Y_1 \propto 
L^{1/\nu}$, with $\nu$ the critical exponent of the correlation length and 
$L$ the lateral system size. To this end we have plotted, in the inset of 
\fig{cumulant}, the cumulant slope $Y_1$ at the critical value 
$\etaprcr=0.9238$ obtained above, versus the system size $L$. By 
performing a fit to the data, the exponent is measured to be $\nu_{\rm 
eff} \approx 0.93$, which is already surprisingly close to the 2D Ising 
value. Obviously, $\nu_{\rm eff}$ must be interpreted as an ``effective'' 
critical exponent.

\subsection{Free energy barrier and interfacial tension}

\begin{figure}
\begin{center}
\includegraphics[width=\figwidth,clip]{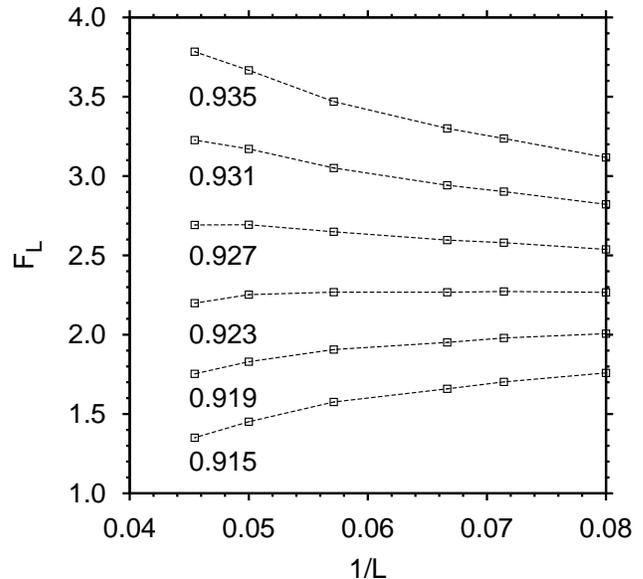}
\caption{\label{barrier} Finite size dependence of the free energy 
barrier $F_L$ between the colloidal vapor and liquid phase, for the 
confined AO model with $q=0.8$ and film thickness $D=5$. Shown is $F_L$ as 
function of $1/L$ at the indicated value of $\etapr$, with $L$ the lateral 
dimension of the simulation box. The barrier was extracted from equal-area 
\cite{muller:1995} distributions $\pc$, see also \fig{prob}.}
\end{center}
\end{figure}

Next, we consider the free energy barrier $F_L$ between the colloidal 
vapor and liquid phase. At the critical point, the grand canonical 
distribution scales with the system size $L$ as \cite{binder:1981, 
nicolaides1988a}
\begin{equation}
\label{eq:sc}
 P^\star_L(\etac) = b_0 L^{\beta/\nu} {\cal P}^0 ( b_0 L^{\beta/\nu} \etac), 
\end{equation}
where $P^\star_L(\etac)$ is the distribution $\pc$ measured in the finite system
at the critical values of $\etapr$ and $\zc$, $b_0$ some non-universal constant,
and ${\cal P}^0$ a function independent of system size (in the present case of
confinement, the critical exponents $\beta$ and $\nu$ assume 2D Ising values).
Recall from \fig{prob} that $F_L$ is given by the peak-to-valley height in the
logarithm of $\pc$. The scaling form of \eq{eq:sc} thus implies that $F_L$
becomes system-size independent at the critical point, providing an alternative
route to locate $\etaprcr$ (see \olcite{potoff:2000} where this approach is
applied to the Lennard-Jones fluid). To locate $\etaprcr$, the barrier is
recorded as function of $L$ for several values of $\etapr$. At the critical
value of $\etapr$, a plateau should be visible. For the confined AO model, the
latter is verified in \fig{barrier}, which shows $F_L$ as function of $1/L$ for
various $\etapr$ around the critical region. The figure shows an increase in
$F_L$ with system size at high $\etapr$, and a decrease at low $\etapr$. The
plateau occurs in the interval $\etapr = 0.923-0.927$. Although not very
precise, this estimate is consistent with the previous result based on the
intersection of the cumulant.

A more precise estimate of $\etaprcr$ may be obtained from the critical 
behavior of the interfacial tension $\gamma_\infty$ in the thermodynamic 
limit. Upon approach of the critical point, starting in the two-phase 
region, the interfacial tension is expected to vanish as \cite{77}
\begin{equation}
 \gamma_\infty = \Gamma_0 t^\mu, \quad \mu = 2-\alpha-\nu,
\end{equation}
with $t=\etapr/\etaprcr-1$ the relative distance from the critical point, 
critical amplitude $\Gamma_0$, and critical exponents listed in 
\tab{tab:prop}. For 2D Ising systems we thus obtain $\mu_{\rm 2D}=1$ and 
for 3D Ising systems $\mu_{\rm 3D}=2\nu$, where in the latter the 
hyperscaling relation $2-\alpha=d\nu$ was used (with $d$ the spatial 
dimension). For 3D bulk, the expected exponent $2\nu$ was already 
confirmed by us in \olcite{68}. In the present case of confinement, 
however, the crossover scaling scenario implies a transition in the 
critical behavior of $\gamma_\infty$ from singular ($\mu=2\nu$) in three 
dimensions to purely regular ($\mu=1$) in two dimensions. This 
particularly affects the {\it slope} of $\gamma_\infty$ versus $t$ at the 
critical point, which should be zero in 3D, and finite positive in 2D.

\begin{figure}
\begin{center}
\includegraphics[width=\figwidth,clip]{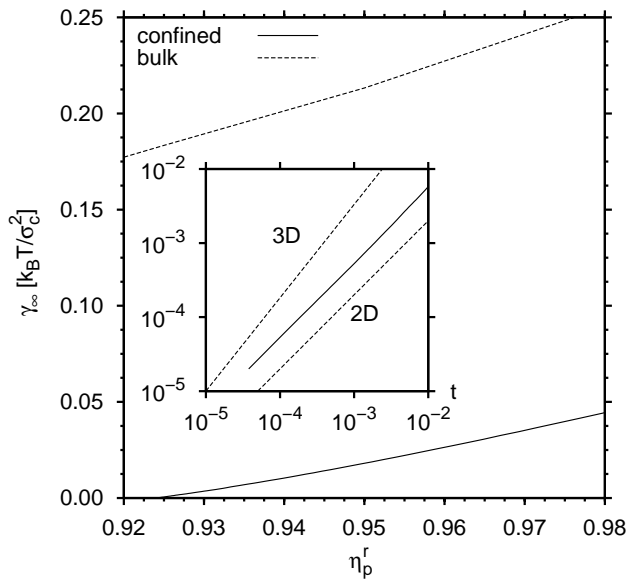}

\caption{\label{tension} {\it main frame:} Interfacial tension 
$\gamma_\infty$ of the confined AO model with $q=0.8$ and $D=5$ as 
function of $\etapr$ (solid curve), as well as the corresponding bulk 
interfacial tension (dashed curve). {\it inset:} $\gamma_\infty$ of the 
confined system as function of the relative distance from the critical 
point $t$, on double logarithmic scales, where $\etaprcr=0.9241$ in $t$ 
was used. The dashed lines illustrate 2D and 3D Ising exponents.}

\end{center}
\end{figure}

In order to test if evidence for this change in critical behavior is 
present in our simulation data, we use the free energy barrier $F_L$ to 
measure the interfacial tension. Following \olcite{binder:1982}, the 
interfacial tension $\gamma_L$ in a confined system of thickness $D$ and 
finite lateral dimension $L \gg D$ equals $\gamma_L = F_L/(2LD)$, where 
the factor of two stems from the use of periodic boundary conditions. The 
thermodynamic limit $L \to \infty$ can be evaluated through an elimination 
of finite size effects using the extrapolation equation \cite{binder:1982}
\begin{equation}
\label{eq:binder}
 \gamma_L = \gamma_\infty +  \frac{a_1}{LD} + \frac{a_2 \ln(L)}{LD},
\end{equation}
where the constants $a_1$ and $a_2$ are {\it a priori} unknown. The interfacial 
tension in the thermodynamic limit $\gamma_\infty(\etapr)$ at a given 
value of $\etapr$ may thus be obtained by fitting \eq{eq:binder} to 
corresponding measurements of $\gamma_L(\etapr)$ in finite systems. We 
have applied this approach using different system sizes between 
$L=12.5-22$ and furthermore assuming $a_2=0$ in \eq{eq:binder}. The latter 
choice is based on empirical findings that the logarithmic term in 
\eq{eq:binder} is typically rather weak, at least for Ising-like systems 
\cite{68, berg:1993}. The result is summarized in \fig{tension}. The main 
frame shows the thermodynamic limit interfacial tension $\gamma_\infty$ as 
function of $\etapr$ for the confined AO model, as well as the bulk 
tension taken from previous work \cite{68}. Note the pronounced {\it 
decrease} in the interfacial tension of the confined system with respect 
to the bulk, a direct consequence of the upward shift in $\etaprcr$. For 
the confined system, the vanishing of $\gamma_\infty$ at the critical 
point yields $\etaprcr=0.9241$ which is consistent with our previous 
estimate based on the intersection of the cumulant. Moreover, the 
interfacial tension seems to vanish with finite slope, a point further 
emphasized in the inset where $\gamma_\infty$ is plotted as function of the
relative distance from the critical point. Note that the interfacial 
tension in the confined system is already well described by the 2D Ising 
exponent.

\subsection{Order parameter}

\begin{figure}
\begin{center}
\includegraphics[width=\figwidth,clip]{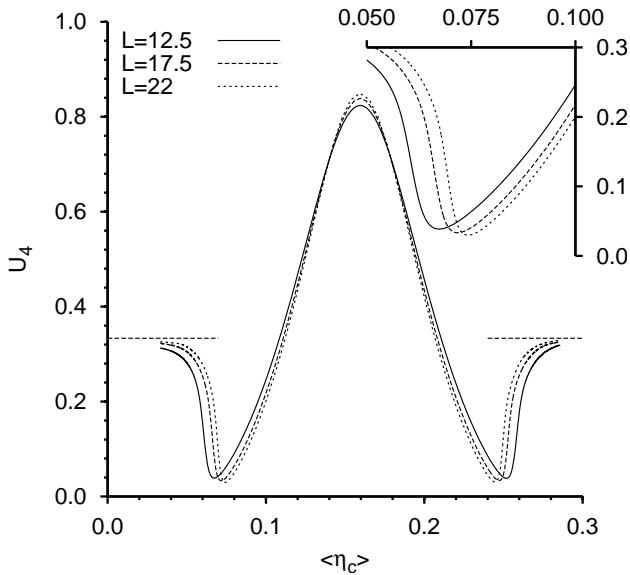}
\caption{\label{qmin} Cumulant ratio $U_4$ as function of the average colloid
packing fraction $\avg{\etac}$. The data were obtained for the confined AO model
with $q=0.8$ at $\etapr=0.93$, film thickness $D=5$, and several lateral
dimensions $L$ as indicated. Dashed horizontal lines correspond to the limiting
value $U_4=1/3$, see details in text. The inset shows the region around
$\etac^-(L,\etapr)$ on an expanded scale.} 
\end{center}
\end{figure}

\begin{figure}
\begin{center}
\includegraphics[width=\figwidth,clip]{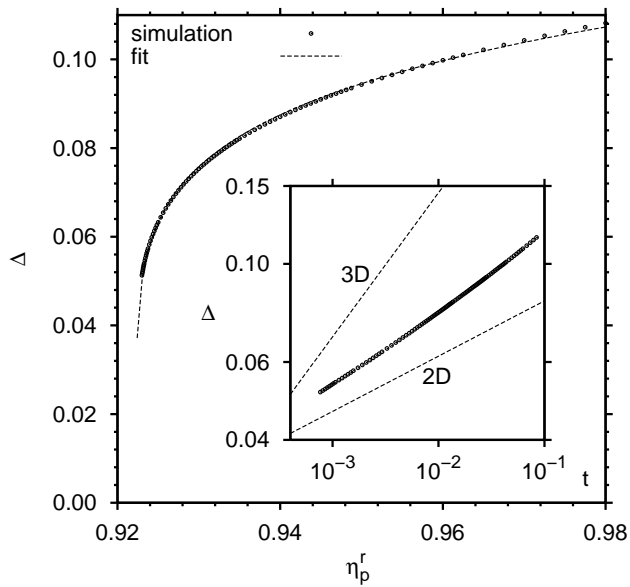}
\caption{\label{order} Order parameter $\Delta$ as function of $\etapr$ for the
confined AO model with $q=0.8$ and film thickness $D=5$. The main frame shows
$\Delta$ in the thermodynamic limit as function of $\etapr$ on linear scales,
where the curve through the simulation data is a fit to \eq{eq7}. The inset
shows the result as function of the relative distance from the critical 
point $t=\etapr/\etaprcr-1$, on double logarithmic scales, where the 
slopes of the lines reflect 2D and 3D Ising critical exponents.}
\end{center}
\end{figure}

Another consequence of the crossover scenario is that the binodal should 
appear ``flatter'', since the critical exponent $\beta$ of the order 
parameter for the 2D Ising model is smaller than in 3D. In this section, 
we use the finite size scaling algorithm of Kim, Fisher, and Luijten 
\cite{kim:2005} to study the critical behavior of the order parameter in 
confinement. The algorithm is based on the dependence of the cumulant 
$U_4$ on the temperature-like variable $\etapr$, the colloid fugacity 
$\zc$, and the system size $L$ (recall that $U_4$ is defined in 
\sect{sec:cum}). For fixed $\etapr$ and $L$, it is straightforward to 
measure $U_4$ and $\avg{\etac}$ as function of $\zc$. A plot of $U_4$ 
versus $\avg{\etac}$, which is thus parameterized by $\zc$, reveals two 
minima separated by a maximum, see \fig{qmin}. The location of the minimum 
at low colloid packing fraction is denoted $\etac^-(L,\etapr)$, with 
$Q^-(L,\etapr)$ the corresponding minimum value. Similarly, the location 
of the minimum at high colloid packing fraction is denoted 
$\etac^+(L,\etapr)$, with $Q^+(L,\etapr)$ the corresponding minimum value. 
Note that the location of the minima, and the corresponding minimum 
values, depend on $\etapr$ and $L$, but obviously not on $\zc$. In the 
thermodynamic limit $L \to \infty$, $U_4$ approaches $1/3$ in the 
one-phase region (horizontal dashed lines in \fig{qmin}). On the 
phase-boundary, $\etac^-(L,\etapr)$ and $\etac^+(L,\etapr)$ approach the 
thermodynamic values $\etacv(\etapr)$ and $\etacl(\etapr)$, respectively, 
while $Q^-(L,\etapr)$ and $Q^+(L,\etapr)$ approach zero 
\cite{kim.fisher:2003}. The $L$-dependence shown in \fig{qmin} is 
consistent with this scenario, see also the inset.

In order for the scaling algorithm to succeed, simulation data for at least
three different system sizes are required. In this work, $L=15$, $17.5$ and $20$
are used. In addition, measurements over a rather broad range in $\etapr$ are
required, starting in the two-phase region and stretching toward the critical
point. Here, five different $\etapr$ are simulated per system size, evenly
distributed over the range $\etapr \approx 0.9-1.0$. Estimates of properties at
intermediate $\etapr$ are obtained using histogram extrapolation
\cite{ferrenberg}. The purpose of the scaling algorithm is to evaluate the order
parameter $\Delta$ as function of $\etapr$ in the thermodynamic limit
\begin{equation}
 \Delta(\etapr) = \lim_{L \to \infty} 
 \frac{ \etac^+(L,\etapr) - \etac^-(L,\etapr) }{2}.
\end{equation}
Following \olcite{kim:2005}, we define the quantities 
\begin{eqnarray}
Q_{\rm min}(L,\etapr) \equiv
  \frac{Q^+(L,\etapr) + Q^-(L,\etapr)}{2}, \\
x(L,\etapr) \equiv Q_{\rm min}(L,\etapr) 
  \ln \left[ \frac{4}{e Q_{\rm min}(L,\etapr)} \right], \\
\label{eq:yy} y(L,\etapr) \equiv 
  \frac{\etac^+(L,\etapr) - \etac^-(L,\etapr)}{2 \Delta(\etapr)}.
\end{eqnarray}
The algorithm starts in the two-phase region, with a value of $\etapr$ 
significantly above its critical value. The peaks in $\pc$ are then well 
separated and the free energy barrier $F_L$ will be large (see for example 
the distribution corresponding to $\etapr=1.0$ in \fig{prob}). This regime 
is called the ``gaussian limit'' because $\pc$ may be described by a sum 
of two gaussians in this case \cite{kim:2005}. In the gaussian limit, it 
can be shown rigorously that the points $(x,y)$ of different system sizes 
$L$, should all collapse onto the line $y = 1+x/2$. Recall that 
$\Delta(\etapr)$ in \eq{eq:yy} is the order parameter in the thermodynamic 
limit at the considered $\etapr$, precisely the quantity of interest, 
which may thus be obtained by fitting until the best collapse onto 
$y=1+x/2$ occurs. Next, $\etapr$ is chosen closer to the critical point, 
the points $(x,y)$ are calculated as before, but this time around 
$\Delta(\etapr)$ is chosen such that the new data set joins smoothly with 
the previous one, yielding an estimate of the order parameter at the new 
$\etapr$. This procedure is repeated as closely as possible to the 
critical point, where $\Delta$ vanishes, yielding an estimate of 
$\etaprcr$.

For the confined AO model, the output of the scaling algorithm is 
illustrated in \fig{order}. The main frame shows the order parameter as 
function of $\etapr$ on linear scales. The dashed curve is a fit to the 
simulation data using \eq{eq7} from which $\etaprcr=0.9223$, $\hat{B}_{\rm 
eff}=0.173$, and $\beta_{\rm eff}=0.17$ are obtained. As before, 
$\beta_{\rm eff}$ plays the role of an effective critical exponent. Note 
that $\beta_{\rm eff}$ is already rather close to the pure 2D Ising value. 
This point is emphasized in the inset of \fig{order}, which shows the 
order parameter as function of the relative distance from the critical 
point $t$, where $\etaprcr = 0.9223$ in $t$ was used. Also included are 
power laws illustrating 2D and 3D Ising critical exponents. As expected, 
the simulation data slowly approach the slope of the 2D Ising exponent. By 
performing additional simulations using larger lateral dimensions $L$, the 
data can in principle be extended to smaller values of $t$, where the pure 
2D Ising exponent will become visible. However, such simulations are 
computationally very demanding and beyond the scope of the present 
investigation. In contrast, adding data at larger values of $t$ in order 
to observe the 3D Ising exponent is not possible, since then we leave the 
critical regime. In hindsight, the thickness $D=5$ considered here is too 
small to observe the full crossover from 3D to 2D Ising critical behavior. 
For such thin films, the critical behavior is essentially 2D Ising. The 
crossover scaling is expected to be visible only in much thicker films, 
where 2D Ising behavior shows up at extremely small $t$.

\begin{figure}
\begin{center}
\includegraphics[width=\figwidth,clip]{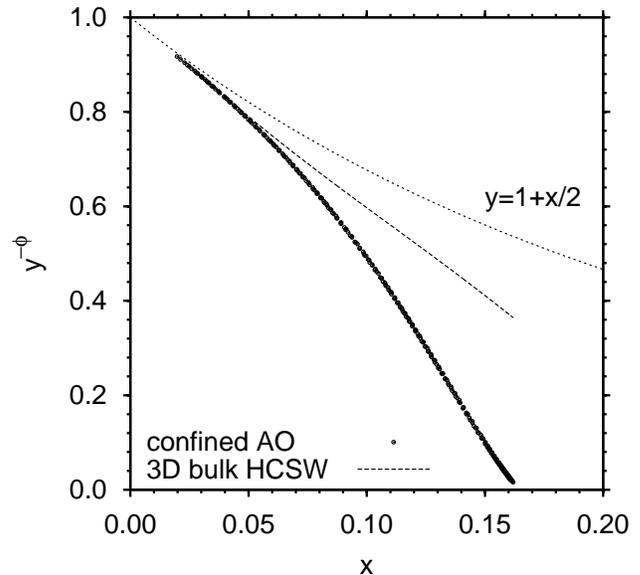}

\caption{\label{scaling} Order parameter scaling curve $y^{-\phi}$, with 
$\phi=1/\beta$ and $\beta=1/8$, for both the 3D bulk HCSW fluid 
\cite{kim:2005}, and the confined AO model ($q=0.8$ and $D=5$) of this 
work. Also shown is the exact small $x$ form $y = 1 + x/2$ of the gaussian 
limit.}

\end{center}
\end{figure}

In addition to the order parameter, the scaling algorithm also yields $y$ as 
function of $x$. The latter function, or scaling curve, is significant 
because it is universal within a universality class. For bulk 3D fluids, 
belonging to the 3D Ising universality class, universality of the scaling 
curve has been verified for the hard-core square-well (HCSW) fluid 
\cite{kim:2005}, the restricted primitive model (RPM) \cite{kim:2005}, the 
decorated lattice gas \cite{kim:2005}, the AO model \cite{verso.vink}, and 
the Widom-Rowlinson mixture \cite{vink:2006}. In the present case of 
confinement, however, the scaling curve is expected to deviate profoundly 
from the bulk 3D Ising form. The latter is verified in \fig{scaling}, which 
shows the scaling curve of the confined AO model obtained in this work, 
together with the scaling curve of the 3D bulk HCSW fluid \cite{kim:2005}. 
Following the convention of \olcite{kim:2005}, the scaling curve has been 
raised to an exponent $-\phi = -1/\beta$, with $\beta=1/8$ the critical 
exponent of the 2D Ising model. An important feature of \fig{scaling} is 
that, for small $x$, the data of the confined AO model correctly approach 
the limiting form $y = 1+x/2$. In addition, we observe that the data from 
the three different system sizes have collapsed accurately onto a single 
curve. As expected, the scaling curve of the 3D bulk HCSW fluid differs 
profoundly from the one of the confined AO model, a direct consequence of 
the different universality classes. Note in particular the large difference 
in $x_c$ at which the scaling curve vanishes. For the HCSW fluid, $x_c 
\approx 0.286$ \cite{kim:2005}, which exceeds the value of the confined AO 
model $x_c \approx 0.165$ by over 60\%. For the 2D Ising model, $x_c \approx 
0.46$ is reported \cite{kim:2005}, which overestimates our value 
significantly. Note, however, that the scaling curve of \fig{scaling} must 
be regarded as an ``effective'' scaling curve, and that deviations from the 
pure 2D Ising form are to be expected.

\subsection{Coexistence diameter}
\label{sec:diam}

Finally, we turn to the critical behavior of the coexistence diameter 
$\delta \equiv (\etacl+\etacv)/2$, in general given by 
\cite{kim.fisher.ea:2003*b}
\begin{equation}\label{eq:coex}
 \delta = \etaccr \left( 1 + A_{2\beta} t^{2\beta}
 + A_{1-\alpha} t^{1-\alpha} + A_1 t \right),
\end{equation}
with $\etaccr$ the colloid packing fraction at the critical point, $t$ the 
relative distance from the critical point, and non-universal amplitudes 
$A_i$. The term proportional to $t^{2\beta}$ is due to pressure mixing, 
and for systems where pressure mixing is absent $A_{2\beta}=0$. It is not 
yet clear which features determine the degree of pressure mixing in a 
fluid. Of the bulk 3D fluids where this issue has been investigated, only 
the RPM exhibits substantial pressure mixing \cite{kim:2005}. In the 
decorated lattice gas, pressure mixing is absent \cite{kim:2005}, and the 
same seems to be the case for the Widom-Rowlinson mixture 
\cite{vink:2006}. Simulations of the HCSW fluid \cite{kim:2005} and the AO 
model \cite{verso.vink} point to rather weak pressure mixing.

To obtain the coexistence diameter of a bulk 3D fluid is still 
challenging. In the present case of confinement the situation is even more 
subtle. Assuming negligible pressure mixing, the critical behavior of the 
diameter is dominated by $t^{1-\alpha}$. The crossover scaling scenario 
then implies a transition from weak singular behavior $\alpha \approx 
0.109$ in 3D, to purely regular behavior $\alpha=0$ in 2D, see 
\tab{tab:prop}. On the other hand, if pressure mixing is important, the 
diameter remains singular and dominated by $t^{2\beta}$, with ultimately 
$\beta=\beta_2$ the critical order parameter exponent of the 2D Ising 
model. To determine which of these scenarios is realized in the confined 
AO model, we again use the finite size scaling approach of Kim, Fisher, 
and Luijten \cite{kim:2005}. Note that, in addition to the order 
parameter, these authors also present an algorithm to extract the 
diameter. As before, the algorithm generates a scaling curve, starting 
with data obtained well away from the critical point, and then recursively 
working its way down toward criticality. However, the quantities needed to 
construct the scaling curve are different. In particular, they involve the 
asymmetry factor
\begin{equation}\label{eq:amin}
 A_{\rm min}(L,\etapr) \equiv \frac{Q^+(L,\etapr) - Q^-(L,\etapr)}
 {Q^+(L,\etapr)+Q^-(L,\etapr)},
\end{equation}
with $Q^\pm(L,\etapr)$ defined previously. Unfortunately, for the confined 
AO model, we were unable to extract the diameter in this way. Closer 
inspection of our data revealed that $A_{\rm min}$ as function of $\etapr$ 
changes sign from negative to positive upon approach of the critical 
point, and this makes the procedure numerically unstable. In contrast, for 
3D bulk systems, $A_{\rm min}$ remains positive (at least for the AO model 
and the Widom-Rowlinson mixture) and so the problem does not occur there.  

\begin{figure}
\begin{center}
\includegraphics[width=\figwidth,clip]{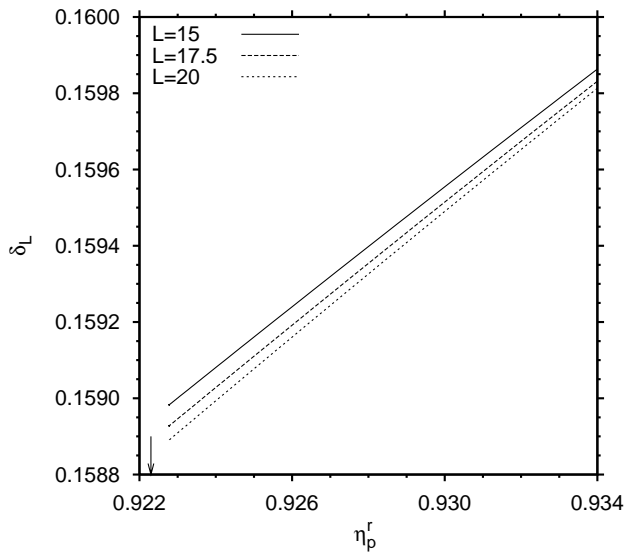}
\caption{\label{diam} Coexistence diameter $\delta_L$ given by \eq{eq:dd} 
of the confined AO model with $q=0.8$ and $D=5$ obtained in finite systems 
of lateral dimension $L$ as indicated. The arrow marks the estimate 
$\etaprcr \approx 0.9223$ obtained from \fig{order}.}
\end{center}
\end{figure}

Hence, our attempt to extract the critical behavior of the diameter in
confinement is unsuccessful. Instead, we follow the more pragmatic approach of
\olcite{brovchenko} and simply show in \fig{diam} the diameter of the finite
system
\begin{equation}\label{eq:dd}
 \delta_L(\etapr) \equiv \frac{ \etac^+(L,\etapr) + \etac^-(L,\etapr) }{2},
\end{equation}
with $\etac^\pm(L,\etapr)$ defined as before. Obviously, these data cannot be
used to extract (effective) critical exponents, but the value $\etaccr \approx
0.159$ of \sect{sec:cum} derived from equal-area distributions $\pc$ seems
confirmed.

\section{Discussion and Conclusion}

\begin{figure}
\begin{center}
\includegraphics[width=\figwidth,clip]{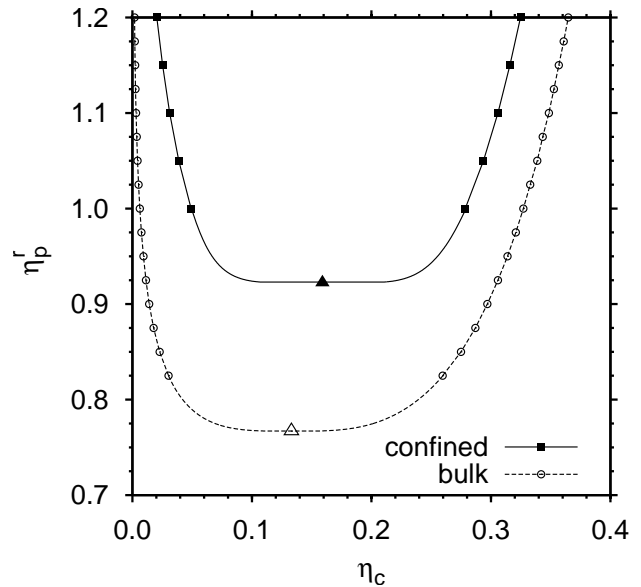}

\caption{\label{coex_curve} Thermodynamic limit binodals, obtained using
finite size scaling, of the AO model with $q=0.8$ in bulk and confinement. 
The dashed curve shows the bulk binodal, the solid curve is the binodal in
the confined system with thickness $D=5$. Triangles mark the corresponding
critical points; squares and circles are raw simulation data obtained in
finite systems away from the critical point.}

\end{center}
\end{figure}

In this work, we have investigated the critical behavior of a 
colloid-polymer mixture confined to a thin film of thickness $D=5$. The 
main finding is that for such thin films, 2D Ising universality is clearly 
visible. The latter is manifested by the critical exponents of the 
correlation length, the interfacial tension, and the order parameter. 
Since the order parameter exponent in 2D Ising systems is smaller than in 
3D, the binodal in the confined system should appear ``flatter''. To 
emphasize this point, we have combined the order parameter data of 
\fig{order} with the coexistence diameter of the largest system in 
\fig{diam}, and constructed the binodal in the thermodynamic limit. 
The result is shown in \fig{coex_curve} as the solid curve, where the 
closed triangle marks the location of the critical point. As expected, 
away from the critical point, the binodal obtained via finite size scaling 
joins smoothly with the raw finite-size simulation data (squares). For 
comparison, the binodal of the bulk system in the thermodynamic limit is 
also shown, taken from previous work \cite{68}. The different curvature of 
the binodals should be detectable in experiments. The result of 
\fig{coex_curve} may furthermore be relevant to Gibbs ensemble simulations 
of fluids in confined geometry. Here, the critical point is typically 
determined via a fitting procedure assuming 3D Ising exponents. In 
contrast, \fig{coex_curve} indicates that for thin films, extrapolations 
using 2D Ising exponents may be more appropriate.

In addition to a flatter binodal, the location of the critical point also 
changes with respect to the bulk. Our results indicate a pronounced shift 
of the critical point toward higher values of $\etapr$, as well as a 
slight increase in the critical colloid packing fraction $\etaccr$. The 
increase in $\etapr$ is consistent with previous work on capillary 
condensation in the AO model that was based on DFT and Gibbs ensemble 
simulations \cite{46}. The behavior of $\etaccr$ is more subtle. For films 
with $D \geq 5 \sigma_c$, DFT shows an {\it increase} in $\etaccr$ with 
respect to the bulk value, while for very thin films a {\it decrease} is 
predicted \cite{46}. It is not obvious if the corresponding Gibbs ensemble 
simulations also follow this trend \cite{46}. At first sight, the increase 
in $\etaccr$ observed in our simulations seems consistent with the trend 
predicted by DFT for films with $D \geq 5 \sigmac$. However, for very 
thick films, $\etaccr$ must ultimately approach its bulk value again, and 
it is not clear how the DFT of \olcite{46} approaches this limit. 
Interestingly, recent Gibbs ensemble simulations of the confined 
Lennard-Jones fluid with film thickness $D=12$ particle diameters, show a 
pronounced {\it decrease} in the critical density \cite{brovchenko}, which 
differs from both the present simulation result, and the DFT trend for 
this rather thick film.

All in all, the critical density seems to depend quite sensitively on the 
details of the particle and wall interactions, as well as on the film 
thickness. Key to a reliable estimate of the critical density is a precise 
description of the coexistence diameter. The latter may be obtained using 
the finite size scaling approach of \olcite{kim:2005}, as was recently 
demonstrated for 3D bulk fluids \cite{kim:2005, vink:2006, verso.vink}. 
The issues pertaining to the shift of the critical density in confinement 
inspired us, in \sect{sec:diam}, to apply this scaling algorithm to the 
confined AO model. Unfortunately, we were unable to extract the diameter 
because the scaling algorithm of \olcite{kim:2005} seems to behave 
profoundly different in confinement; the source is a numerical instability 
arising from a change in sign of the asymmetry factor given by 
\eq{eq:amin}. At this point, we see no reliable way to extract the 
coexistence diameter of the confined AO model; nor do we understand the 
significance of the change in sign in $A_{\rm min}$. To resolve these 
issues would be the subject of further work.

Regarding the order parameter, no difficulties were encountered in applying
\olcite{kim:2005} to the confined AO model. This was demonstrated by the
accurate collapse of the data from different system sizes onto a single scaling
curve. Also the estimated effective critical exponent $\beta_{\rm eff}$ is
consistent with the crossover scaling scenario. Nevertheless, the large
discrepancy in $x_c$ at which the scaling curve vanishes, between the confined
AO model and the 2D Ising model, is concerning. To understand the source of this
discrepancy, too, would require further work.

\acknowledgments

We are grateful to the {\it Deutsche Forschungsgemeinschaft} (DFG) for support
(TR6/A5 and TR6/D3). JH also acknowledges support of the DFG under Grants No. HO
2231/2-1 and HO 2231/2-2.

\end{document}